# Physisorption of DNA bases on finite-size nanoribbons from graphene, phosphorene, and silicene: Insights from density functional theory

Mukesh Tumbapo[1], Matthew B. Henry[1], Sanjiv K. Jha[2], and Benjamin O. Tayo[1]

[1]*Department of Engineering and Physics, University of Central Oklahoma Edmond, OK 73034, USA*

[2]*Department of Physics, East Central University, Ada OK 74820, USA*

Authors to whom correspondence should be addressed: sjha@ecok.edu, btayo@uco.edu

## ABSTRACT

The ability to detect and discriminate DNA bases by reading it directly using simple and cost-effective methods is an important problem whose solution can produce significant value for areas such as cancer and human genetic disorders. Two-dimensional (2D) materials have emerged as revolutionary materials for electronic DNA sequencing with strong potentials for fast, single-nucleotide direct-read DNA sequencing with a minimum amount of consumables. Among 2D materials, graphene is the most explored for DNA sequencing. This is due to its commercial availability. The major hindrance of graphene is its hydrophobicity, which causes DNA bases to stick to its surface, slowing down translocation speed, and making single-base discrimination difficult as multiple bases interact with graphene at any given time. It is therefore essential that other elemental 2D materials beyond graphene be investigated. Using density functional theory (DFT), we studied the electronic interaction of DNA bases physisorped onto the surface of nanoribbons from graphene, phosphorene, and silicene. By comparing the change in energy band gap, binding energy and density of states (DOS), we observe that phosphorene performs better than graphene and silicene for DNA sequencing using the physisorption modality.

*Keywords*: DNA sequencing; Nanoribbons; Nucleobases, 2D materials; Nanomaterials; DFT.

## I. INTRODUCTION

Among the large family of 2D materials, the scientific community has taken great interest in the possibilities of graphene for DNA sequencing as evidenced by theoretical investigations[1-3] as well as three almost simultaneous experimental demonstrations of DNA translocating through graphene pores.[4-6] Recently, technologies have been developed to create graphene nanogaps in a cost-effective manner and connect them to high frequency electronic waveguides for rapid sequencing.[7] One advantage of graphene is its commercial availability, and it can easily be patterned down to the desired sensing element using techniques such as AFM and TEM.[4-8] In addition, the single-layer nature of graphene is comparable to the distance between adjacent bases of a single-stranded DNA (ssDNA) molecule, hence it is ideal for single-base resolution. The major issue with graphene is its hydrophobicity, which causes DNA bases to stick to its surface,[9-11] reducing translocation speed and increasing error rates.[8,12,13] Furthermore, pristine graphene is a gapless semiconductor, hence it is unsuitable for electronic DNA sequencing using modalities such as tunneling current modulations or FETs.[12]

With all the disadvantages of graphene, it is essential that other elemental 2D materials beyond graphene[14-16] be explored for DNA sequencing. Two such candidate materials are phosphorene[14,17] and silicene.[15,18] These materials have desirable properties such as high carrier mobility and tunable direct band gaps,[14,15,18-20] which makes them suitable for electronic DNA sequencing applications. Moreover, phosphorene is hydrophilic and biocompatible, hence ideal for sequencing applications.[21,22] A recent periodic density functional theory (DFT) study demonstrated electronic DNA sequencing using phosphorene nanoribbons.[22,23] Using DFT, we studied the electronic interaction of DNA bases physisorped onto the surface of graphene, phosphorene, and silicene nanoribbons. By comparing the change in energy band gap, binding energy and DOS, we observe that phosphorene performs better than graphene and silicene for DNA sequencing using the physisorption modality.[8]

## II. MATERIALS AND COMPUTATIONAL METHODS

Our model is based on detecting variations in the in-plane current[8,12,23] in a nanoribbon (width ~ 1.0 – 2.0 nm) due to physisorption of DNA bases onto the surface of the 2D nanoribbon as shown in Fig. 1. Such a dimension is necessary for single-base resolution since the distance between bases in ssDNA is ~ 0.63 nm[24] and the largest base, guanine, has a size of ~ 0.754 nm. A nucleotide consist of a sugar, phosphate group and base, and the phosphate group and sugar are the same for all nucleotides. Hence, in our idealized model, we shall consider only the interaction of the four DNA bases (guanine (G), adenine (A), cytosine (C) and thymine (T)) with 2D nanoribbons.[12,25] For simplicity, we will refer to our nanoribbon systems using the following abbreviations: GNR (graphene nanoribbon), PNR (phosphorene nanoribbon) and SNR (silicene nanoribbon).

For each 2D material, finite-size nanoribbons (width ~ 1.0 – 2.0 nm) were created. Hydrogen passivation[12,23,26] was used to terminate the dangling bonds at the edges. Structural minimization was performed for the individual bases and membranes separately, and then for the combined system of nanoribbon and bases. All bases were placed 3.0 Å above surface of nanoribbon before geometry optimization. The structural relaxation calculations were performed at the B3LYP level of theory using the 6-31G (d, p) basis set, with a force convergence cutoff of 0.02 eV/Å.[23] All calculations were performed using the GAUSSIAN 16 software package.[27] Computational resources were provided by the University of Central Oklahoma Buddy Supercomputing Center.[28] To assess the potential for electronic base detection, three evaluation parameters were computed, namely, the energy band gap, binding energy, and DOS. The equations for the energy gap, binding energy, and DOS can be found in Ref. 29. As graphene is the most widely studied 2D material for DNA sequencing, the results for phosphorene and silicene were benchmarked against graphene.

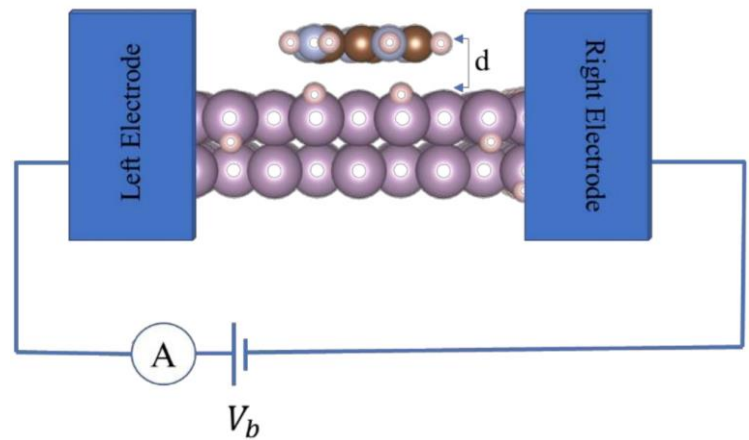

**FIG.1**. Schematic of nanoribbon sensing device. Changes in electronic current due to physisorption of DNA bases onto the surface of a 2D nanoribbon can be detected. The in-plane current flows from left to right, that is perpendicular to the DNA backbone during DNA traversal.

## III. RESULTS

### A. Graphene Nanoribbon (GNR)
Our GNR system has 9 dimer lines (zigzag direction), a width of ~1.00 nm along the zigzag direction and a length of ~1.17 nm along the armchair direction. Table 1 shows the band gap and binding energies for each system.

**TABLE I**. Energy gap and binding energy for GNR systems.

| System | $E_{gap}$ (eV) | $E_{bind}$ (eV) |
|---|---|---|
| **GNR + G** | 0.260 | -0.592 |
| **GNR + A** | 0.257 | -0.546 |
| **GNR + C** | 0.262 | -0.578 |
| **GNR + T** | 0.258 | -0.423 |

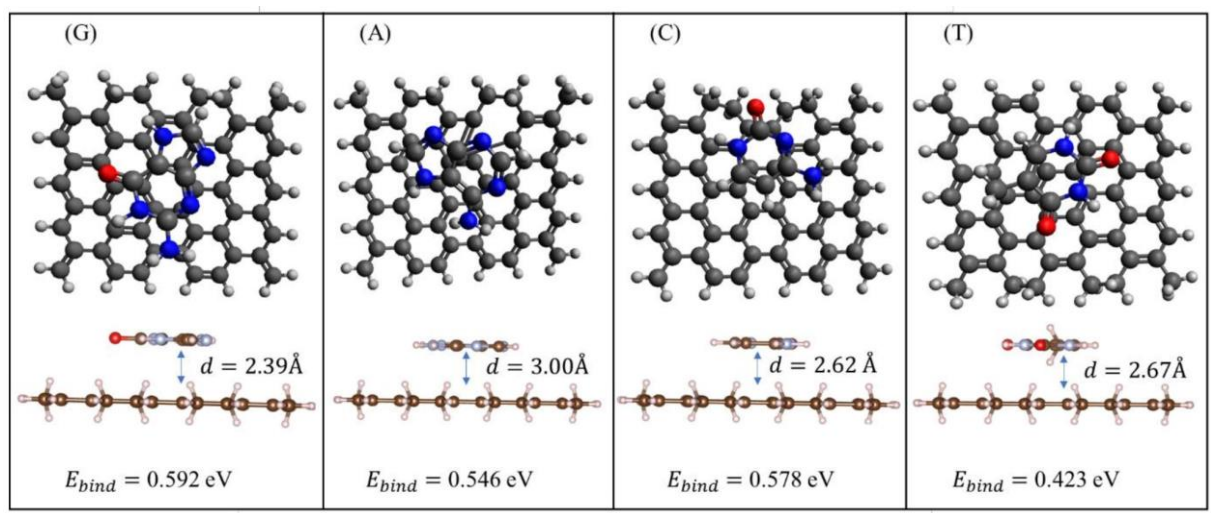

**FIG. 2**. Relaxed structures of GNR + nucleobases. The absolute value of computed binding energies and equilibrium adsorption heights are shown.

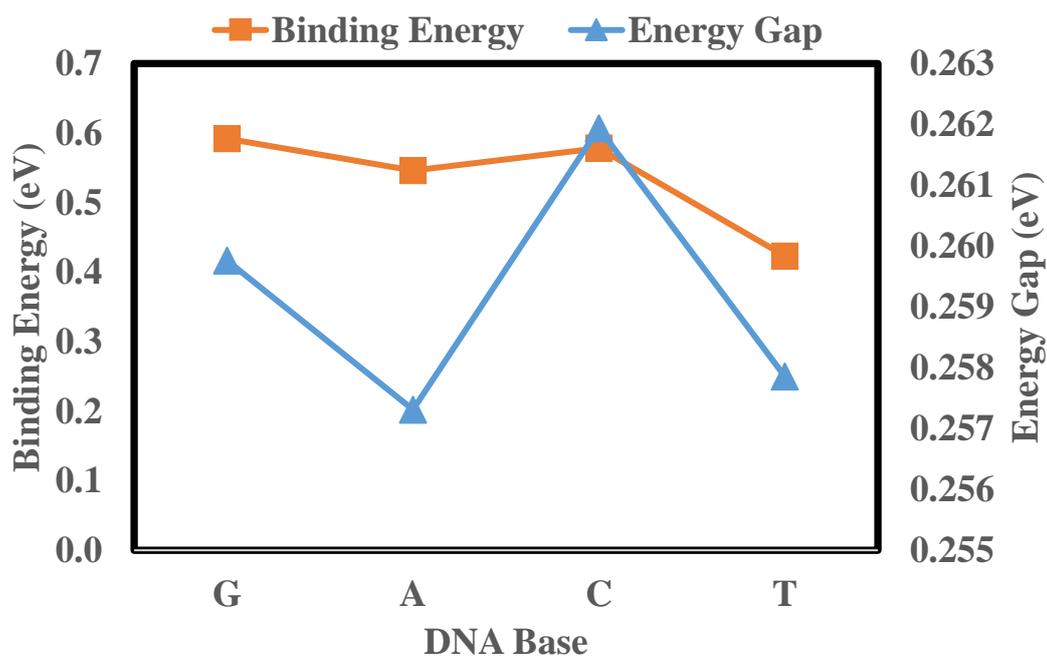

**FIG. 3**. Binding Energy and Energy Gap of GNR with each DNA base placed on top of GNR.

The band gap of pristine GNR is 0.259 eV. Table 1 shows that the modulation in the band gap of GNR due to adsorption of DNA bases is negligible. The largest change in band gap is produced by base C

(0.003 eV). For binding energy, the largest value is produced by base G (0.592 eV). The optimized geometry for each GNR system is shown in Fig. 2. The equilibrium adsorption height is smallest in base G (2.39 Å) and largest in A (3.00 Å) and follows the trend G < C < T < A. The variation of binding energy and energy gap for each base is shown in Fig. 3.

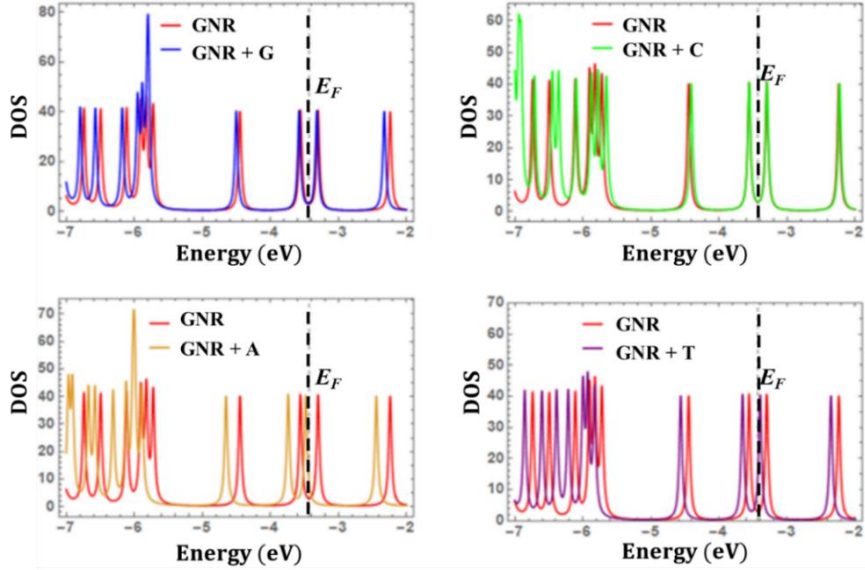

**FIG. 4**. DOS of GNR and GNR + base systems. $E_F$ is the Fermi energy of pristine GNR.

Figure 4 shows that the modulation in DOS for pristine GNR due to interaction with DNA bases is small. The small variation in energy band gap and DOS for GNR is undesirable, as this can give rise to low tunneling current signal-to-noise ratios.[8,12] Also, the relatively large binding energy of DNA bases suggest the potential for bases to stick to the surface of the graphene nanoribbon, which can produce major issues with sequencing such as low translocation speed and high error rates.[9-11]

**B. Phosphorene Nanoribbon (PNR)**

Our PNR system has 9 dimer lines (zigzag direction), a width of ~1.33 nm along the zigzag direction and a length of ~1.32 nm along the armchair direction. Table 2 shows the band gap and binding energies for each system.

**TABLE II**. Energy gap and binding energy for PNR systems.

| System | $E_{gap}$ (eV) | $E_{bind}$ (eV) |
|---|---|---|
| **PNR + G** | 2.680 | -0.330 |
| **PNR + A** | 2.783 | -0.293 |
| **PNR + C** | 3.022 | -0.182 |
| **PNR + T** | 3.055 | -0.169 |

The calculated band gap for PNR is 3.038 eV. Unlike GNR, the band gap modulations for PNR are larger. The largest change is produced by base G (reduction in band gap of 0.358 eV). For binding energy, the largest value is produced by base G (0.330 eV). The binding energies follow the trend G > A > C > T, which is consistent with periodic DFT results for nanoribbons from phosphorene,[23] graphene,[30] and $MoS_2$.[12] Figure 5 shows the optimized geometry for each base. The equilibrium adsorption height is smallest in base G (2.636 Å) and largest in C (3.171 Å) and follows the trend G < T < A < C. The variation of binding energy and energy gap for each base is shown in Fig. 6, which is consistent with previous results using nanoribbons from phosphorene,[23] graphene,[30] and $MoS_2$.[12] Figure

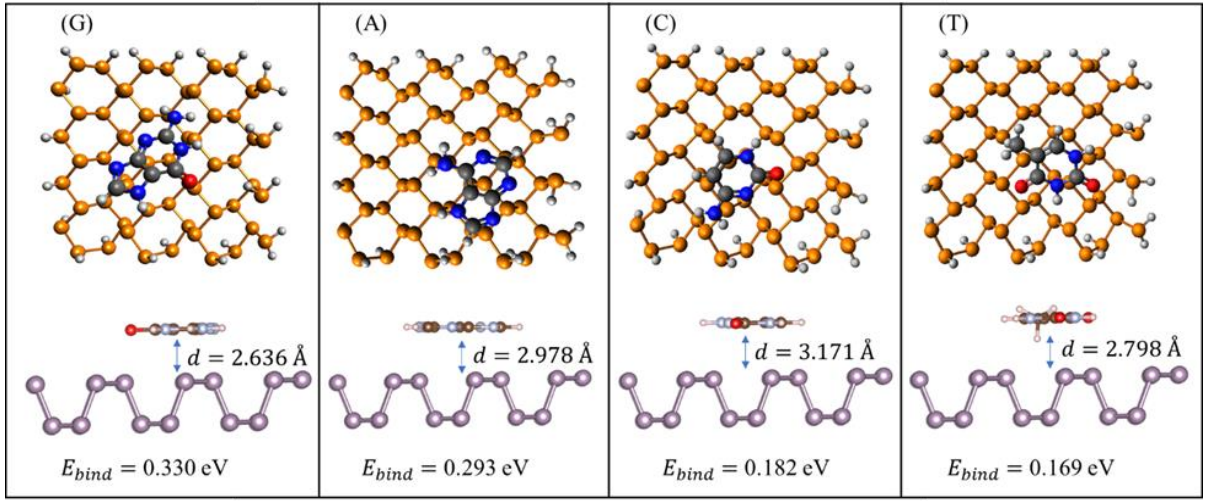

**FIG. 5**. Relaxed structures of PNR + nucleobases. The absolute value of computed binding energies and equilibrium adsorption heights are shown.

7 shows the DOS for PNR with DNA bases. The large modulation in band gap energy and DOS, and smaller binding energies with DNA bases makes PNR a superior material to GNR for electronic DNA sequencing using the physisorption modality.[8]

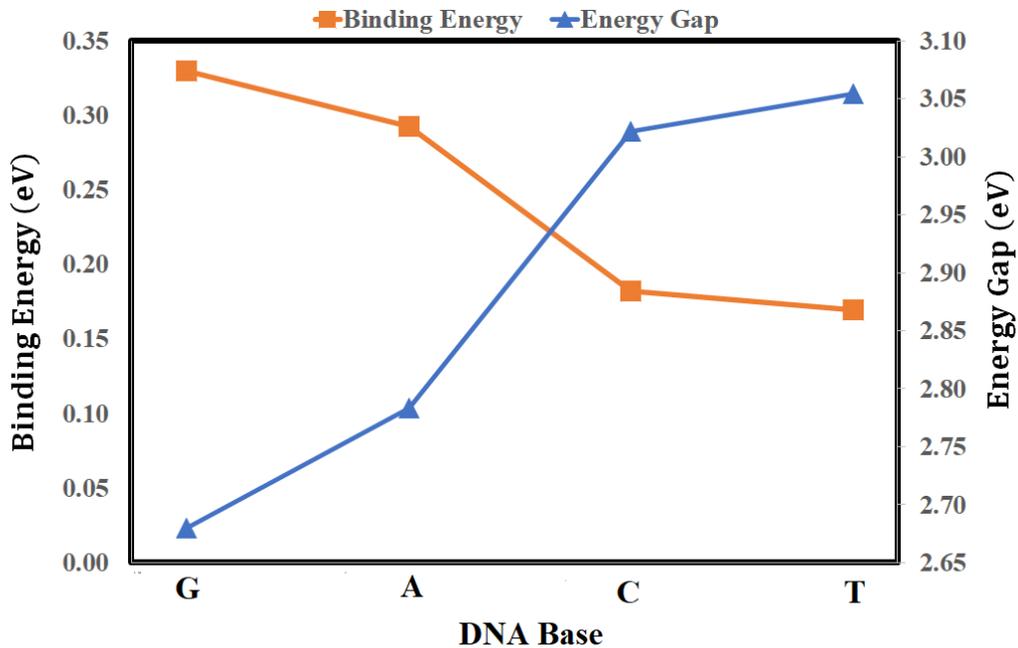

**FIG. 6**. Binding Energy and Energy Gap of PNR with each DNA base placed on top of PNR. Adapted from Henry *et al.*, AIP Advances **11**, 035324 (2021).

### C. Silicene Nanoribbon (SNR)

Our SNR system has 11 dimer lines (zigzag direction), a width of ~2.18 nm along the zigzag direction and a length of ~2.07 nm along the armchair direction. Table 3 shows the band gap and binding energies for each system. The calculated band gap for the SNR system is 0.348 eV. The largest change in band gap is observed in base A (an increase of 0.024 eV). For binding energy, the largest value is produced by base C (0.633 eV).

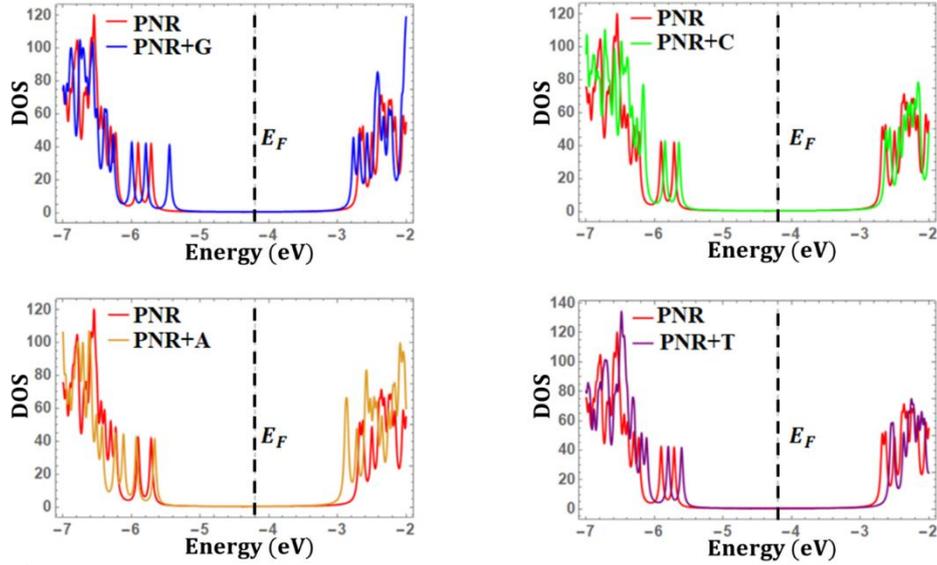

**FIG. 7**. DOS of PNR and PNR + base systems. $E_F$ is the Fermi energy of pristine PNR.

**TABLE III**. Energy gap and binding energy for SNR systems.

| System | $E_{gap}$ (eV) | $E_{bind}$ (eV) |
|---|---|---|
| **SNR + G** | 0.329 | -0.493 |
| **SNR + A** | 0.372 | -0.180 |
| **SNR + C** | 0.366 | -0.633 |
| **SNR + T** | 0.344 | -0.094 |

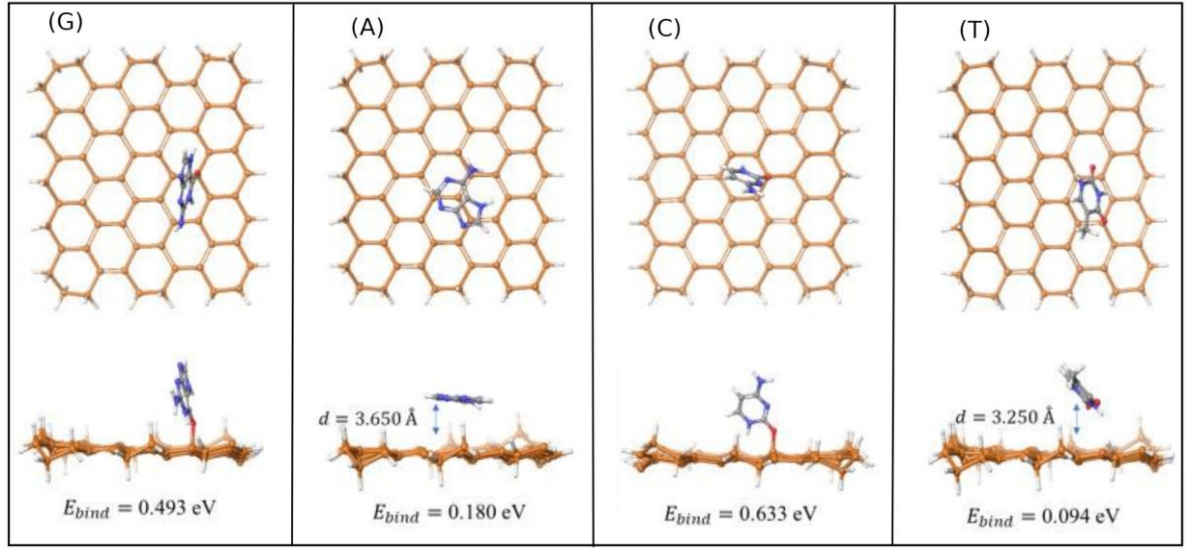

**FIG. 8**. Relaxed structures of SNR + nucleobases. The absolute value of computed binding energies and equilibrium adsorption heights are shown.

Figure 8 shows the optimized geometry for SNR and DNA bases. We noticed that bases G and C chemisorbed onto the SNR, reacting with the surface and forming a covalent bond. No sticking was observed for A and T. The optimized vertical distance above the surface of the SNR was calculated as

3.650 Å and 3.250 Å for SNR+A and SNR+T, respectively. This sticking was observed even when the size (length and width) of the nanoribbon sheet was increased.

D. Comparing GNR, PNR, and SNR

We studied the interaction of DNA bases with small-width nanoribbons from various 2D materials. Due to the difference in C-C bond length in graphene and P-P bond length in phosphorene, the widths of the nanoribbons are 1.00 nm (GNR) and 1.33 nm (PNR). For SNR, the width is 2.18 nm since there are 11 dimer lines, compared to 9 for GNR and PNR. We observe from Table 4 that binding energies of PNR are lower than GNR. For SNR, the large binding energies for G and C are due to the formation of covalent bonds with the surface of the silicene nanoribbon. Table 5 shows the change in band gap for the different materials. It shows that the variations in energy gap for PNR and SNR are larger compared to GNR.

TABLE IV. Energy Gap and Binding Energy for GNR, PNR, and SNR.

| Base | Energy Gap (eV) | | | Binding Energy (eV) | | |
|---|---|---|---|---|---|---|
| | GNR | PNR | SNR | GNR | PNR | SNR |
| Pristine | 0.259 | 3.038 | 0.348 | - | - | - |
| G | 0.260 | 2.680 | 0.329 | 0.592 | 0.330 | 0.493 |
| A | 0.257 | 2.783 | 0.372 | 0.546 | 0.293 | 0.180 |
| C | 0.262 | 3.022 | 0.366 | 0.578 | 0.182 | 0.633 |
| T | 0.258 | 3.055 | 0.344 | 0.423 | 0.169 | 0.094 |

TABLE V. Change in energy gap for GNR, PNR, and SNR systems due to the presence of DNA base.

| | $\Delta E_{gap}$ (eV) | | |
|---|---|---|---|
| Base | GNR | PNR | SNR |
| G | +0.001 | -0.358 | -0.019 |
| A | -0.002 | -0.255 | +0.024 |
| C | +0.003 | -0.016 | +0.018 |
| T | -0.001 | +0.017 | -0.004 |

Finally, Fig. 9 shows the binding energy comparison for the different materials. The small binding energies of DNA bases in PNR suggest minimal sticking of DNA bases. Hence, hydrophobic interactions are expected to be minimal in PNR compared to graphene.[9-11]

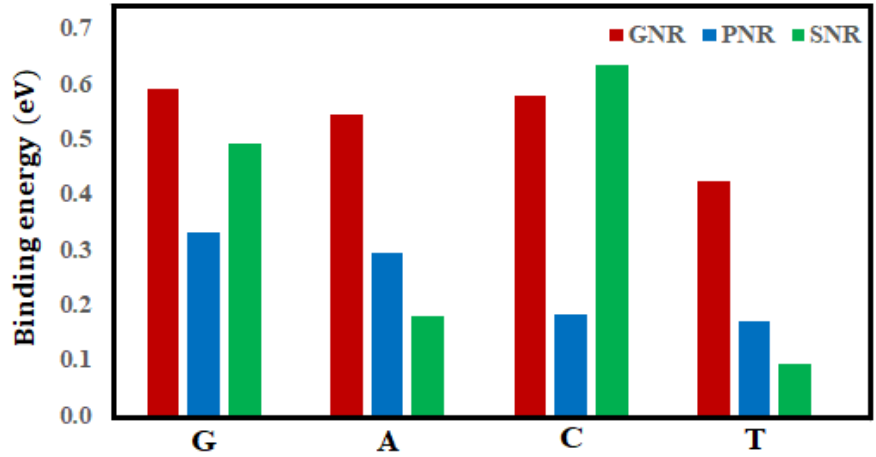

**FIG. 9**. Comparison of binding energies for GNR, PNR, and SNR.

## IV. DISCUSSION

In our idealized model, the effect of ions and solvating water molecules were not taken into account.[12,22,23,25] Ions and solution are expected to alter the magnitude of band gaps and binding energies, but not the trends.[31,32] Only initial configurations in which the bases were planar to the surface of the nanoribbons were treated. Different orientations would slightly alter the band gaps and binding energies, but not the trends.[29] Even though our simplified study was carried out using local basis sets, our findings agree nicely with results using periodic calculations, and seems to capture van der Waals effects.[23]

## V. CONCLUSION

In summary, using DFT, we studied the interaction of DNA basis with finite-size nanoribbons from graphene, phosphorene, and silicene. We observe that binding energies of DNA bases using nanoribbons from phosphorene (0.1 - 0.4 eV) are smaller compared to silicene (0.1 - 0.7 eV) and graphene (0.4 - 0.6 eV). This shows that minimal sticking of DNA bases to nanoribbon surface is expected for phosphorene devices. The binding energy sequence was determined as G > C > A > T for graphene, G > A > C > T for phosphorene, and A > G ~ C > T for silicene. Bases C and G were observed to form covalent bonds with the silicene nanoribbon. Furthermore, nanoribbons from phosphorene and graphene show a characteristic change in the density of states for each base. The change in band gap due to interaction with DNA bases is more significant for phosphorene ($\Delta E_{gap} \sim 16 - 358$ meV) compared to silicene ($\Delta E_{gap} \sim 4 - 24$ meV) and graphene ($\Delta E_{gap} \sim 1 - 3$ meV). Unlike in graphene and phosphorene, bases G and C chemisorbed onto the silicene nanoribbon surface, reacting with the surface and forming a covalent bond. Based on these preliminary results, we conclude that for DNA sequencing using physisorption, phosphorene performs better than graphene and silicene. The formation of covalent bonds between DNA bases and silicene makes silicene to be non-ideal, as bases are expected to stick to its surface during DNA translocation.

## AUTHOR'S CONTRIBUTIONS

All authors contributed equally to this work.

## ACKNOWLEDGEMENT

This work was funded by the Faculty On-Campus Grant and the RCSA Grant from the Office of Research and Sponsored Programs at the University of Central Oklahoma (UCO). B. Tayo would like to thank the College of Mathematics and Science at UCO for the CURE-STEM research funds.

## DATA AVAILABILITY

The data that support the findings of this study are available from the corresponding author upon reasonable request.